\documentclass[a4paper]{article}
\usepackage{graphicx,amsmath,amssymb}
\newcommand{\haak}[1]{\left(#1\right)}
\newcommand{\rhaak}[1]{\left [#1\right]}
\newcommand{\lhaak}[1]{\left | #1\right |}
\newcommand{\ahaak}[1]{\left\{#1\right\}}
\newcommand{\gem}[1]{\left\langle #1\right\rangle}

\newcommand{\haakl}[1]{\left(#1\right.}
\newcommand{\haakr}[1]{\left.#1\right)}

\newcommand{\lhaakl}[1]{\left |#1\right.}

\newcommand{\floor}[1]{\left\lfloor #1\right\rfloor}
\newcommand{\half}{\frac{1}{2}}

\renewcommand{\imath}{\text{i}}

\begin{document}
\title{Exact conjectured expressions for correlations in the dense O$(1)$ loop 
model on cylinders} 
\author{Saibal Mitra and Bernard Nienhuis\\
Instituut voor Theoretische Fysica\\
Universiteit van Amsterdam\\
1018 XE Amsterdam
The Netherlands}
\date{\today}
\maketitle
\begin{abstract}
We present conjectured exact expressions for two types of 
correlations in the dense O$(n=1)$ 
loop model on $L\times \infty$ square lattices with periodic boundary 
conditions. These are the probability 
that a point is surrounded by $m$ loops and the probability that $k$ consecutive 
points on a row are on the same or on
different loops. The dense O$(n=1)$ loop model is equivalent to the bond percolation model at the critical point. The former probability can be interpreted in terms of the bond percolation problem as giving the probability that a vertex is on a cluster that is surrounded by $\floor{m/2}$ clusters and $\floor{(m+1)/2}$ dual clusters. The conjectured expression for this probability involves a binomial determinant that is known to give weighted enumerations of cyclically symmetric plane partitions and also of certain types of families of nonintersecting lattice paths. By applying Coulomb gas methods to the dense O$(n=1)$ loop model, we obtain new conjectures for the asymptotics of this binomial determinant.
\end{abstract} 

\section{Introduction}
\begin{figure}[t]
\begin{center}
\includegraphics[width=0.5\textwidth]{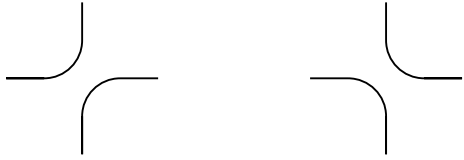}
\caption{The two vertices of the dense O(1) loop model}\label{fig:vrt}
\end{center}
\end{figure}

In this article we consider the dense $O$(1) loop model \cite{loop} defined on an 
$L\times\infty$ square lattice with 
periodic boundary conditions. The geometry of the lattice is thus a cylinder 
with circumference $L$. 
The dense $O(n)$ loop model can be defined as follows. At each 
vertex the four edges meeting there 
are connected with equal probability in either of the two ways shown in Fig.\ 
\ref{fig:vrt}.

\begin{figure}[h]
\begin{center}
\includegraphics[width=0.5\textwidth]{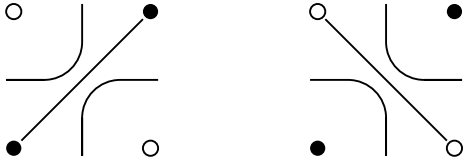}
\end{center}
\caption{The mapping of a loop configuration to a bond configuration of the corresponding bond percolation problem. A bond is either put on an edge of the square lattice formed by the 
$\bullet$ or on the dual edge 
orthogonal to it on the square lattice formed by the $\circ$.}\label{fig:perc}
\end{figure}

If $L$ is even, the set of edges connected to each other form closed loops with probability 1. If 
$L$ is odd there is a loop spanning the entire length of the cylinder.
To a state consisting of $l$ closed paths (loops) a weight of $n^{l}$ is
assigned. In this article we consider exclusively the case $n=1$, in which all configurations have equal weight. At this point the model can be mapped to the bond percolation problem at criticality, the six vertex model and the XXZ-spin chain at $\Delta=-\half$ (see \cite{baxkelwu}). The corresponding bond percolation problem is defined on one of the sublattices of the dual lattice. The states of the
dense O(1) loop model are in direct bijection with bond configurations of the bond percolation problem, see Fig.\ \ref{fig:perc}. 
\setlength{\unitlength}{\textwidth}

\begin{figure}[ht]
\begin{center}
\includegraphics[width=0.5\textwidth]{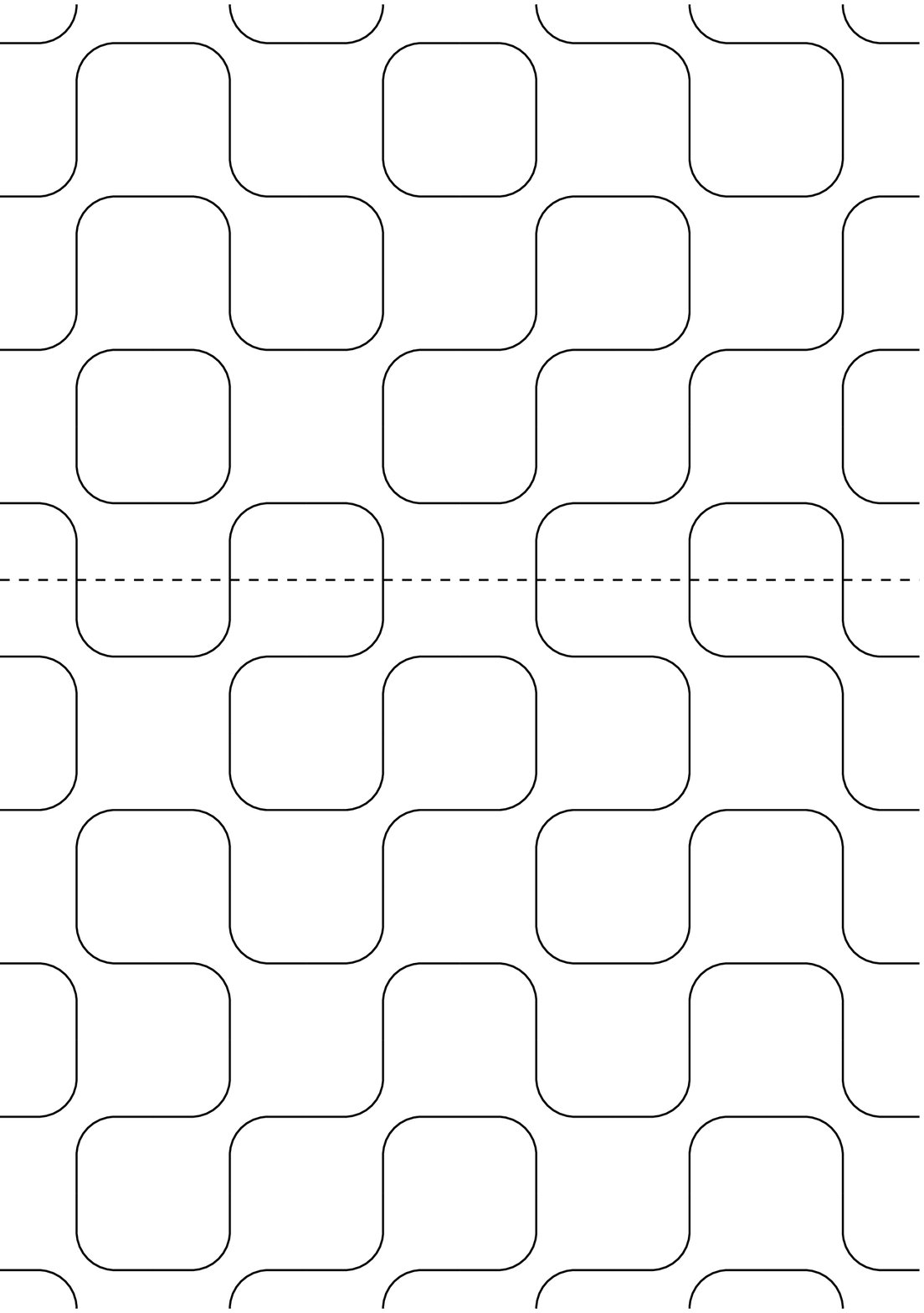}
\caption{Part of a typical configuration of the O$(1)$ loop model on a
$6\times \infty$ cylinder, periodic in the horizontal direction and
extending to infinity in both vertical directions.
The connectivity state of the half cylinder below the row indicated by the dashed line is $\haak{}\haak{}\haak{}$,
above the row it is $\haakr{}\haak{}\haakl{}\haak{}$.}\label{fig:O1}
\end{center}
\end{figure}

Recently, a number of conjectures have been obtained about correlations in this model. These involve probabilities that loops intersect 
a horizontal cut between vertices (henceforth referred to as a row) 
in certain prescribed ways. A row intersects loops at a total of $L$ points. We define the connectivity state of the row as the way these $L$ points are connected to each other by the loops via the half cylinder below the row. See Fig.\ \ref{fig:O1} for an example. Connectivity states can be denoted by strings of parentheses \cite{mtr}. If a point at position $i$ is connected to a point at position $j$, then this is represented by a parenthesis at position $i$ matching with a parenthesis at position $j$. Because of periodic boundary conditions, a right parenthesis at position $i$ can match with a left parenthesis at position $j$ also if $i<j$. For odd $L$, one point will not be connected to any other points on the row. That point is on an unpaired line. This can be denoted by a line $\lhaakl{}$. For example, the connectivity state $\haakr{}\lhaakl{}\haakl{}\haak{}$, denotes the state in which the first point is connected to third point over the boundary, the second point is not connected to any other point, and the fourth point is connected to the fifth point.

Several conjectures about the properties of this probability distribution have been obtained, see \cite{BdGN1,Strog2,Strog3,degier02112852, mtr}. These results can be interpreted as conjectures for probabilities of certain events on the top row of a half infinite cylinder. A recent summary of these results is given in \cite{degier02112852}. Surprising connections with the enumeration of certain symmetry classes of alternating sign matrices \cite{rob} have been discovered. In this article we will make use of the following observations: If the sum of 
all probabilities is normalized to the number of $L\times L$ half turn symmetric alternating sign matrices, denoted as $A_{\text{HT}}\haak{L}$, then all probabilities are conjectured to be integers. It is conjectured that the probability of the connectivity state $\haak{\haak{\haak{\ldots}}}$ for even $L$ and $\lhaakl{}\haak{\haak{\haak{\ldots}}}$ for odd $L$ with all the parenthesis nested inside each other is 1 in this normalization. For even $L$, it has been proven \cite{kup2} that:
\begin{equation}\label{ahtev}
A_{\text{HT}}\haak{L}=2\prod_{k=1}^{L/2-1}\frac{3}{4}\frac{\haak{3k+2}!\haak{3k-
1}!k!\haak{k-1}!}{\haak{2k+1}!
^{2}\haak{2k-1}!^{2}}
\end{equation}
For odd $L$ the number of $L\times L$ half turn symmetric alternating sign matrices is conjectured to be 
\cite{rob}:
\begin{equation}\label{ahtodd}
A_{\text{HT}}\haak{L}=\prod_{k=1}^{\haak{L-
1}/2}\frac{4}{3}\frac{\haak{3k}!^{2}k!^{2}}{\haak{2k}!^{4}}
\end{equation}
In some expressions for correlations in this paper the total number of $ n\times n $ alternating sign matrices, $A(n)$ appears. This is given by \cite{zeil,kup1}:
\begin{equation}\label{al}
A\haak{n}=\prod_{k=0}^{n-1}\frac{\haak{3k+1}!}{\haak{n+k}!}
\end{equation}

A convenient way to compute probabilities of events at a row on the complete cylinder is to consider separately the connectivity states on both half spaces below and above the row. Since events above and below the row are independent, the probability distribution over the connectivity states contains all the information needed.
\section{Probability of being surrounded by $m$ loops}
In \cite{mitradmtcs} we have presented a conjectured exact formula for the probability $P\haak{L,m}$ that a 
face of the lattice on a cylinder of circumference $L$ is surrounded by $m$ loops. This conjecture is only 
valid for even $L$. Here we will explain how we obtained this conjecture and also generalize this conjecture to odd $L$.

Since it is conjectured that all probabilities of connectivity states are integers when the sum of all probabilities is normalized to be $A_{\text{HT}}\haak{L}$, multiplying $P\haak{L,m}$ by $A_{\text{HT}}\haak{L}^{2}$ should be an integer. Let's define $Q\haak{L,m}\equiv P\haak{L,m} 
A_{\text{HT}}\haak{L}^{2}$. The values for $Q\haak{L,m}$ that we used to guess this function are given in table \ref{qlm}.
\begin{table}
\begin{center}
\caption{Values for $Q\haak{L,m}$}\label{qlm}
\begin{tabular}{|c|r|r|r|r|r|r|}\hline
$m$ & \multicolumn{6}{c|}{$L$}\\\hline
&\multicolumn{1}{c|}{2}&\multicolumn{1}{c|}{4}&\multicolumn{1}{c|}{6}&\multicolumn{1}{c|}{8}&\multicolumn{1}{c|}{10}&\multicolumn{1}{c|}{12}\\\hline\hline
0&3& 70& 13167& 20048886& 247358122583& 24736951705389664\\\hline
1&1& 29& 6081& 9938153& 129127963303& 13446619579882992\\\hline
2&0& 1& 351& 744189& 11416135802& 1338566241796974\\\hline
3&0& 0& 1& 4707& 112088578& 17605459620761\\\hline
4&0& 0& 0& 1& 66197& 18952950999\\\hline
5&0& 0& 0& 0& 1& 956385\\\hline
6&0& 0& 0& 0& 0& 1\\\hline
\end{tabular}
\end{center}
\end{table}

\begin{table}
\begin{center}
\caption{Values for $C_{L/2-m}\haak{L}$}\label{clm}
\begin{tabular}{|c|r|r|r|r|r|r|}\hline
$m$& \multicolumn{6}{c|}{$L$}\\\hline
&\multicolumn{1}{c|}2&\multicolumn{1}{c|}{4}&\multicolumn{1}{c|}{6}&\multicolumn{1}{c|}{8}&
\multicolumn{1}{c|}{10}&\multicolumn{1}{c|}{12}\\\hline\hline
0&3& 72& 13869& 21537270& 270190791369& 27414197906689626\\\hline
1&1& 29& 6084& 9952274& 129464229047& 13499435968309125\\\hline
2&0& 1& 351& 744193& 11416400590& 1338642053600985\\\hline
3&0& 0& 1& 4707& 112088583& 17605464402686\\\hline
4&0& 0& 0& 1& 66197& 18952951005\\\hline
5&0& 0& 0& 0& 1& 956385\\\hline
6&0& 0& 0& 0& 0& 1\\\hline
\end{tabular}
\end{center}
\end{table}
Guessing integer sequences is not difficult when the integers factorize into small primes, see e.g.\ \cite{krattdet,degier02112852,mtr}. In such cases simple product formulae can often be found by inspecting the prime factorizations of the integers. In this case, however, some integers have large prime factors. By inspecting the numbers $Q\haak{L,m}$, we noted that they are close to certain coefficients of the characteristic polynomial of the Pascal matrix.
We denote by $C_{p}\haak{L}$ the absolute values of the coefficients of the 
characteristic polynomial of the 
$L\times L$ Pascal matrix:
\begin{equation}\label{pasc}
\det_{1\leq r,s\leq L}\rhaak{\binom{r+s-2}{r-1}-
x\delta_{r,s}}=\sum_{n=0}^{L}C_{n}\haak{L}\haak{-x}^{n}
\end{equation}
It turns out that $Q\haak{L,m}$ is approximately $C_{L/2-m}\haak{L}$ for $m<L/2 -1$. If $m\geq L/2 -1$ then the two quantities are exactly the same:
\begin{equation}
\begin{split}
Q\haak{L,L/2}&=C_{0}\haak{L}=1\\
Q\haak{L,L/2-1}&=C_{1}\haak{L}
\end{split}
\end{equation}
In table \ref{clm} we have listed the values of $C_{L/2-m}\haak{L}$.
By comparing the entries of tables \ref{qlm} and \ref{clm} we see that
\begin{equation}
Q\haak{L,L/2-2}=C_{2}\haak{L}-L/2
\end{equation}
It was a little more difficult to obtain this guess for $Q\haak{L,L/2-3}$:
\begin{equation}
Q\haak{L,L/2-3}=C_{3}\haak{L}-\haak{L/2 -1}C_{1}\haak{L}
\end{equation}
The results obtained so far led us to the idea that $Q\haak{L/2 - k}$ might be a linear combination of $C_{k-2r}\haak{L}$ for $r=0\ldots \floor{k/2}$, where the coefficients are polynomials in $L$. Using this idea we obtained:
\begin{equation}
\begin{split}
Q\haak{L,L/2-4}&=C_{4}\haak{L}-\haak{L/2 -2}C_{2}\haak{L}+\frac{L}{4}\haak{\frac{L}{2}-3}\\
Q\haak{L,L/2-5}&=C_{5}\haak{L}-\haak{L/2 -3}C_{3}\haak{L}+\half\haak{\frac{L}{2}-4}\haak{\frac{L}{2}-1}C_{1}\haak{L}
\end{split}
\end{equation}

At this point we were starting to run out of data. The expression for $Q\haak{L,L/2-5}$ had been guessed using only $Q\haak{10,0}$ and $Q\haak{12,1}$. Only $Q\haak{12,0}$ was left to guess $Q\haak{L,L/2-6}$. When we substitute $L/2\rightarrow m +k$ in the polynomials multiplying $C_{r}\haak{L}$ in the expressions for 
$Q\haak{L,L/2-k}$, we see that the polynomials depend only on $m$. The results obtained so far led us to the conjecture:
\begin{equation}\label{guess}
Q\haak{L,m}=\sum_{r=0}^{L/4-m/2} K_{r}\haak{m}C_{L/2-m-2 r}\haak{L}
\end{equation}
where
\begin{equation}\label{kr}
\begin{split}
K_{0}\haak{m}=&1\\
K_{1}\haak{m}=&-\haak{m+2}\\
K_{2}\haak{m}=&\half\haak{m+1}\haak{m+4}
\end{split}
\end{equation}
Using the value for $Q\haak{12,0}$, we can see that $K_{3}\haak{0}$ should be $-2$. Since $K_{1}\haak{0}=-2$ and $K_{2}\haak{0}=2$, we guessed that $K_{r}\haak{0}=2\haak{-1}^{r}$ for $r>0$. Assuming that this pattern holds and that $K_{r}\haak{m}$ factors into linear terms, we guessed that the sequence of polynomials could be $-\haak{m+2}$, $1/2!\haak{m+1}\haak{m+4}$, $-1/3!\haak{m+1}\haak{m+2}\haak{m+6}$, $1/4!\haak{m+1}\haak{m+2}\haak{m+3}\haak{m+8}$, etc.
We thus put our faith in the formula:
\begin{equation}\label{plm}
Q\haak{L,m}=C_{L/2-m}\haak{L}+\sum_{r=1}^{L/4-m/2}(-1)^{r}C_{L/2-m-
2r}\haak{L}\frac{m+2r}{m+r}\binom{m+r}{r}
\end{equation}

Besides fitting the data, any guess for $Q\haak{L,m}$ has to be consistent with the normalization:
\begin{equation}\label{nrm1}
\sum_{m=0}^{L/2}Q\haak{L,m}=A_{\text{HT}}\haak{L}^{2}
\end{equation}
Also, we had noted that putting a factor $\haak{-1}^{m}$ in this summation yields the total number of $ L\times L $ alternating sign matrices:
\begin{equation}\label{nrm2}
\sum_{m=0}^{L/2}\haak{-1}^{m}Q\haak{L,m}=A\haak{L}
\end{equation}
We have proved that these two relations follow from \eqref{plm}. See the next section for more examples of such identities. Also we have generated data for $L=14$ for $Q\haak{L,m}$ and verified that the formula predicts the correct values.

For odd $L$ it turns out that $Q\haak{L,m}$ is close to $C_{\haak{L-1}/2-m}\haak{L}$. In this case we were led to this conjecture:
\begin{equation}\label{plmodd}
Q\haak{L,m}=\sum_{r=0}^{\frac{\haak{L-1}}{4}-\frac{m}{2}}(-1)^{r}\rhaak{C_{\frac{\haak{L-1}}{2}-m-2r}\haak{L}-C_{\frac{\haak{L-1}}{2}-m-2r-1}
\haak{L}}\binom{m+r}{r}
\end{equation}

The probabilities $P\haak{L,m}$ can be interpreted in terms of the related bond percolation problem, see Fig.\ \ref{fig:perc}. Loops are boundaries of percolation clusters, separating clusters from dual clusters. If a point is surrounded by no loops then that point is on a cluster that wraps round the cylinder. A point surrounded by $k$ loops is on a cluster that is surrounded by $\floor{\haak{k+1}/2}$ dual clusters and $\floor{k/2}$ clusters.

\section{Arbitrary weights for loops winding round a point}
Using the conjectured expression for $P\haak{L,m}$, we can give arbitrary weights to loops winding round a point. By $\phi\haak{L,z,x}$ we denote the operator that gives loops that 
wind round the point $x$ a weight of $z$. By translational invariance its expectation value is independent of $x$, and in the following we will suppress the argument $x$. We can write:
\begin{equation}\label{phidef}
\gem{\phi\haak{L,z}}=\sum_{m=0}^{L/2}P\haak{L,m}z^{m}
\end{equation}
Conjecture \eqref{plm} and \eqref{plmodd} imply that:
\begin{equation}\label{normq2}
\begin{split}
\gem{\phi\haak{L,z=2\cos\haak{\theta}}}&=\frac{D\haak{L,\theta}}{A_{\text{HT}}\haak{L}^{2}}\text{ 
if } L\text{ is even,}\\
\gem{\phi\haak{L,z=2\cos\haak{\theta}}}&=\frac{1}{2\cos\haak{\theta/2}}\frac{D\haak{
L,\theta}}{A_{\text{HT}}\haak{
L}^{2}}\text{ if } L\text{ is odd}
\end{split}
\end{equation}
Here
\begin{equation}\label{dltheta}
D\haak{L,\theta}=\exp\haak{-\imath\theta L/2}\det_{1\leq r,s \leq 
L}\rhaak{\binom{r+s-2}{r-1}+\exp\haak{\imath\theta}\delta_{r,s}}
\end{equation}

The proofs of these relations can be obtained by substituting the conjectures for $P\haak{L,m}$ in \eqref{phidef} and performing the binomial summations. We show here how this works for even $L$. For odd $L$ the proof is similar. For even $L$, \eqref{plm} implies:
\begin{equation}\label{qbk}
\sum_{m=0}^{L/2}Q\haak{L,m}z^{m}=\sum_{k=0}^{L/2}B_{k}\haak{z}C_{L/2-k}\haak{L}
\end{equation}
where 
\begin{equation}\label{bk}
\begin{split}
B_{0}\haak{z}&=1\\
B_{k}\haak{z}&=\sum_{r=0}^{k/2}\haak{-1}^{r}\frac{k}{k-r}\binom{k-r}{r}z^{k-2 r}\text{ for }k\geq 1
\end{split}
\end{equation}
The summand in this equation, which we name $f\haak{k,r,z}$, satisfies the recurrence:
\begin{equation}
f\haak{k+2,r+1,z}-z f\haak{k+1,r+1,z}+f\haak{k,r,z}=0
\end{equation}
Summing this recurrence over $r$ from $0$ to $k/2$ yields:
\begin{equation}\label{rec}
B_{k+2}\haak{z}-z B_{k+1}\haak{z}+B_{k}\haak{z}=0
\end{equation}
From \eqref{bk} we see that $B_{1}\haak{z}=z$ and $B_{2}\haak{z}=z^2-2$. The solution of \eqref{rec} satisfying these two initial conditions is
\begin{equation}
B_{k}\haak{z}=2\cos\haak{k\theta}
\end{equation}
where $2\cos\haak{\theta}=z$. Inserting this in \eqref{qbk} and using the relation $C_{L/2+k}=C_{L/2-k}$ \cite{lunnon} yields \eqref{normq2} for even $L$.

Only for $\theta$ a multiple of $\pi/3$ are exact evaluations of 
$D\haak{L,\theta}$ known \cite{lozenge,andrews,krattdet}. For $\theta=0$ it is known that the determinant is given as $A_{\text{HT}}\haak{2L}/A\haak{L}$. We have found that for $\theta$ any multiple of $\pi/3$, $D\haak{L,\theta}$ can be expressed in terms of $A\haak{L}$ and $A_{\text{HT}}\haak{L}$. These expressions are given in Table \ref{eval}. To prove these formulas, one can divide the product expressions for these determinants given in \cite{lozenge, krattdet} by the expressions given in Table \ref{eval}, substituting the formulas for $A\haak{L}$ and $A_{\text{HT}}\haak{L}$ given by \eqref{ahtev}, \eqref{ahtodd} and \eqref{al}. The resulting expressions then become rational functions which are easily proven to be identically 1.
\begin{table}
\begin{center}
\caption{Values for $D\haak{\theta}$ for $\theta$ a multiple of $\pi/3$. Except 
for the first entry of the 
Table, the expression in terms of $A\haak{L}$ and $A_{\text{HT}}\haak{L}$ is 
new.}\label{eval}
\begin{tabular}{|c|c|c|}\hline
$\theta$ & \multicolumn{2}{c|}{$D\haak{L,\theta}$}\\\hline
&even $L$&odd $L$\\\hline
0 & $\frac{A_{\text{HT}}\haak{2L}}{A\haak{L}}$ 
&$\frac{A_{\text{HT}}\haak{2L}}{A\haak{L}}$\\\hline
$\pi/3$ & $A_{\text{HT}}\haak{L}^{2}$ 
&$\sqrt{3}A_{\text{HT}}\haak{L}^{2}$\\\hline
$2 \pi/3$ & $A\haak{L}$ & $A\haak{L}$\\\hline
$\pi$ & $A\haak{\frac{L}{2}}^{4}$&0\\\hline
\end{tabular}
\end{center}
\end{table}

\section{Asymptotic behavior of $\gem{\phi\haak{L,z}}$}
In this section we will obtain the asymptotic behavior of 
$\gem{\phi\haak{L,z}}$ by applying Coulomb gas techniques \cite{nienhuis,nienhuis2} to the dense O$(1)$ loop model. First one randomly assigns an orientation to the loops of a state. One then assigns a Boltzmann weight to the oriented loops so that the sum of the Boltzmann weights of the two orientations for each loop is unity. This is achieved by assigning a weight of $\exp\haak{\imath\pi/12}$ each time a loop takes a right turn and $\exp\haak{-\imath\pi/12}$ each time a left turn is taken. To a configuration of oriented loops a height function $h\haak{r}$ can be defined on the faces of the lattice as follows. Rotate all the arrows at the boundaries between the faces 90 degrees anti-clockwise. The rotated arrows define the (discrete) gradient of the height function. The difference between two neighboring faces is $\pm 1$. It is assumed that in the scaling limit the discrete height function may be replaced by a real valued function with a Gaussian measure $\mathcal{M}\rhaak{\ahaak{h\haak{r}}}\sim\exp\haak{-H_{\text{g}}\haak{h}}$, where
\begin{equation}
H_{\text{g}}=\frac{\pi}{6}\int d^{2}x\haak{\nabla h}^{2}
\end{equation}
is the Hamiltonian of the Gaussian model.
By identifying operators in the loop model with operators in the Gaussian model, the asymptotic behavior of correlations can be inferred by computing appropriate Gaussian averages.

Unfortunately, there is no simple representation of $\phi\haak{L,z}$ in terms of the field $h\haak{r}$. However, there exists a related operator, defined on the infinite plane, that can be 
represented as a two point function. For two points $x_{1}$ and $x_{2}$ in the plane, consider the operator $\phi\haak{z,x_{1},x_{2}}$ that gives loops winding round $x_{1}$ or $x_{2}$, but not both, a weight of $z$. All other loops have a weight of 1. Interpreting $\phi\haak{L,z}$ as a one point function, the exponents associated with the large $L$ behavior of $\gem{\phi\haak{L,z}}$ should be half those of $\gem{\phi\haak{z,x_{1},x_{2}}}$ describing the large $\lhaak{x_{1}-x_{2}}$ behavior. The operator $\phi\haak{z,x_{1},x_{2}}$ can be expressed as:
\begin{equation}\label{phizx}
\phi\haak{z,x_{1},x_{2}}=\exp\haak{\imath\haak{\theta+\pi/3}h\haak{x_{1}}}\times\exp\haak{-\imath\haak{\theta-\pi/3}h\haak{x_{2}}}
\end{equation}
Here $\theta=\arccos\haak{z/2}$. An oriented loop winding round $x_{1}$ or $x_{2}$, but not both points, is given a phase factor of $\exp\haak{\pm\imath\theta}$ by including this phase factor and by dividing by $\exp\haak{\pm\imath\pi/3}$, which is the phase factor for any loop in the loop model. Loops winding round both $x_{1}$ and $x_{2}$ pick up a phase factor of $\exp\haak{\pm 2\imath\pi/3}$. This has the effect of changing the sign of the phase of such loops and thus leaves the weight unchanged.

It follows from \cite{nijs} that the exponents $r_{n}$ associated with $\phi\haak{z,x_{1},x_{2}}$ are given by:
\begin{equation}
r_{n}=\frac{3}{2}\frac{\haak{\theta+2\pi 
n}^2}{\pi ^{2}}-\frac{1}{6}
\end{equation}

We thus expect that:
\begin{equation}\label{asymp}
\gem{\phi\haak{L,z}}=\sum_{n=-\infty}^{\infty}A_{n}\haak{L,\theta} L^{-
\frac{3}{4}\frac{\haak{\theta+2\pi 
n}^2}{\pi ^{2}}+\frac{1}{12}}
\end{equation}
Where the coefficients $A_{n}\haak{L,\theta}$ are analytical corrections and have the expansion:
\begin{equation}
A_{n}\haak{L,\theta}=a_{n,0}\haak{\theta}+\sum_{k=2}^{\infty}\frac{a_{n,k}\haak{
\theta}}{L^{k}}
\end{equation}
The leading asymptotic behavior is thus:
\begin{equation}
\gem{\phi\haak{L,z}}\sim a_{0,0}\haak{\theta} L^{-
\frac{3\theta^{2}}{4\pi^{2}}+\frac{1}{12}}
\end{equation}
Assuming that the $a_{n,k}\haak{\theta}$ are analytical functions, it follows from \eqref{asymp} that these functions for fixed $k$ are analytical continuations of each other:
\begin{equation}
a_{n,k}\haak{\theta}=a_{0,k}\haak{\theta+2\pi n}
\end{equation}
It also follows from \eqref{asymp} that:
\begin{equation}
a_{-n,k}\haak{\theta}=a_{n,k}\haak{-\theta}
\end{equation}

\begin{figure}
\begin{center}
\includegraphics[width=0.5\textwidth]{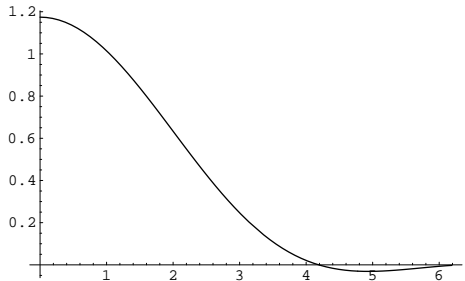}
\caption{The function $a_{0,0}\haak{\theta}$.}\label{graph}
\end{center}
\end{figure}

By calculating the asymptotic expansions of the exact evaluations of $D\haak{L,\theta}$ for $\theta$ a multiple of $\pi/3$ given in Table \ref{eval}, it is possible to obtain exact values for function $a_{0,0}\haak{\theta}$ at those values of $\theta$ for both even 
and odd $L$. In appendix \ref{ap:barnes} we present a simple method to obtain the expansions. In Table \ref{amp} we have listed the exact values for function $a_{0,0}\haak{\theta}$. We have also numerically computed this function for many values in the interval $0\leq\theta\leq 2\pi$. A graph of $a_{0,0}\haak{\theta}$ is shown in figure \ref{graph}.
\begin{table}
\begin{center}
\caption{Exact values of the amplitude function 
$a_{0,0}\haak{\theta}$.}\label{amp}
\begin{tabular}{|c|c|c|}\hline
$\theta$ & \multicolumn{2}{c|}{$a_{0,0}\haak{\theta}$}\\\hline
&even $L$&odd $L$\\\hline
$0$& $ 2^{\frac{1}{9}}\haak{\frac{3}{8}}^{\frac{1}{12}}\exp\haak{-\zeta'\haak{-
1}}$
&$\half\sqrt{3}2^{\frac{1}{9}}\haak{\frac{3}{8}}^{\frac{1}{12}}\exp\haak{-
\zeta'\haak{-1}}$\\\hline
$\pi/3$& $ 1$&$1$\\\hline
$2\pi/3$&$ 
2^{\frac{1}{9}}\haak{\frac{3}{8}}^{\frac{1}{12}}\frac{\Gamma\haak{\frac{2}{3}}}
{\Gamma\haak{\frac{1}{3}}}\exp\haak{-\zeta'\haak{-1}} $&
$\sqrt{3}2^{\frac{1}{9}}\haak{\frac{3}{8}}^{\frac{1}{12}}\frac{\Gamma\haak{\frac
{2}{3}}}
{\Gamma\haak{\frac{1}{3}}}\exp\haak{-\zeta'\haak{-1}}$\\\hline
$\pi$& $2^{-
\frac{1}{3}}\haak{\frac{\Gamma\haak{\frac{2}{3}}}{\Gamma\haak{\frac{1}{3}}}}^{2}
$&? \\\hline
$4\pi/3$ & $0$&$0$\\\hline
\end{tabular}
\end{center}
\end{table}
We can see from Table \ref{amp} that for odd $L$ the amplitude for $\theta$ a multiple of $\pi/3$ except $\pi$ is $\frac{\sqrt{3}}{2\cos\haak{\theta/2}}$ times the amplitude for even $L$. Numerical computations appear to indicate that this relation holds for all $\theta$. If this is indeed the case then it follows from \eqref{normq2} and \eqref{hlf} that the determinant $D\haak{L,\theta}$ has the same asymptotic behavior for even and odd $L$.  

The function $D\haak{L,\pi}$ is zero for odd $L$. The limit 
\[\tilde{\phi}\haak{L}\equiv\lim_{\theta\rightarrow\pi}\frac{D\haak{L,\theta}}{2 \cos\haak{\theta/2}A_{\text{HT}}\haak{L}^{2}}\] is, however, nonzero and finite. The divergence of the amplitude function at $\theta=\pi$ suggests the presence of the logarithmic term $\log\haak{L}$ in the large $L$ behavior of $\tilde{\phi}\haak{L}$. According to \eqref{asymp} all exponents should be $2/3$ plus an integer. We thus expect that the large $L$ behavior is given by an expansion of the form:
\begin{equation}
\tilde{\phi}\haak{L}=\haak{\sum_{k=0}^{\infty}a_{k}L^{-\frac{2}{3}-k}}+\haak{\sum_{k=0}^{\infty}b_{k}L^{-\frac{2}{3}-k}}\log\haak{L}
\end{equation}
where $a_{1}=b_{1}=0$.
We have numerically verified this expansion.

\section{Enumerations of families of nonintersecting lattice paths and cyclically symmetric plane partitions}
In \cite{lozenge} it is shown that the determinant
\begin{equation}\label{dlth}
\det_{1\leq r,s\leq L}\rhaak{\binom{r+s-2}{r-1}+\exp\haak{\imath\theta}\delta_{r,s}}
\end{equation}
in \eqref{dltheta} gives a weighted enumeration of certain types of nonintersecting lattice paths and cyclically symmetrical plane partitions. Consider a family of nonintersecting lattice paths on an $L\times L$ square lattice subjected to the constraints:
\begin{enumerate}
\item A path is allowed to move one step to the right or downward.
\item A path is required to start from a point $\haak{0,k}$ with $0\leq k\leq L-1$
\item A path starting from $\haak{0,k}$ is required to end at the point $\haak{k,0}$
\end{enumerate}
See Fig.\ \ref{fig:path} for an example.
\setlength{\unitlength}{0.8\textwidth}
\begin{figure}
\begin{center}
\begin{picture}(0.6,0.6)(-0.08,-0.08)
\put(0,0){\makebox(0,0){$\bullet$}}
\put(0.2,0){\makebox(0,0){$\bullet$}}
\put(0.4,0){\makebox(0,0){$\bullet$}}
\put(0.6,0){\makebox(0,0){$\bullet$}}

\put(0,0.2){\makebox(0,0){$\bullet$}}
\put(0.2,0.2){\makebox(0,0){$\bullet$}}
\put(0.4,0.2){\makebox(0,0){$\bullet$}}
\put(0.6,0.2){\makebox(0,0){$\bullet$}}

\put(0,0.4){\makebox(0,0){$\bullet$}}
\put(0.2,0.4){\makebox(0,0){$\bullet$}}
\put(0.4,0.4){\makebox(0,0){$\bullet$}}
\put(0.6,0.4){\makebox(0,0){$\bullet$}}

\put(0,0.6){\makebox(0,0){$\bullet$}}
\put(0.2,0.6){\makebox(0,0){$\bullet$}}
\put(0.4,0.6){\makebox(0,0){$\bullet$}}
\put(0.6,0.6){\makebox(0,0){$\bullet$}}

\put(-0.05,0){\makebox(0,0){$A_{4}$}}
\put(-0.05,0.2){\makebox(0,0){$A_{3}$}}
\put(-0.05,0.4){\makebox(0,0){$A_{2}$}}
\put(-0.05,0.6){\makebox(0,0){$A_{1}$}}
\put(0,-0.05){\makebox(0,0){$E_{4}$}}
\put(0.2,-0.05){\makebox(0,0){$E_{3}$}}
\put(0.4,-0.05){\makebox(0,0){$E_{2}$}}
\put(0.6,-0.05){\makebox(0,0){$E_{1}$}}
\put(0,0.6){\line(1,0){0.2}}
\put(0.2,0.6){\line(0,-1){0.2}}
\put(0.2,0.4){\line(1,0){0.2}}
\put(0.4,0.4){\line(0,-1){0.2}}
\put(0.4,0.2){\line(1,0){0.2}}
\put(0.6,0.2){\line(0,-1){0.2}}
\put(0,0.4){\line(0,-1){0.2}}
\put(0,0.2){\line(1,0){0.2}}
\put(0.2,0.2){\line(0,-1){0.2}}
\put(0.2,0){\line(1,0){0.2}}
\end{picture}
\caption{A typical configuration of nonintersecting lattice paths enumerated by the determinant in \eqref{dlth} for $L=4$. Paths starting at $A_{i}$ have to end at $E_{i}$.}\label{fig:path}
\end{center}
\end{figure}
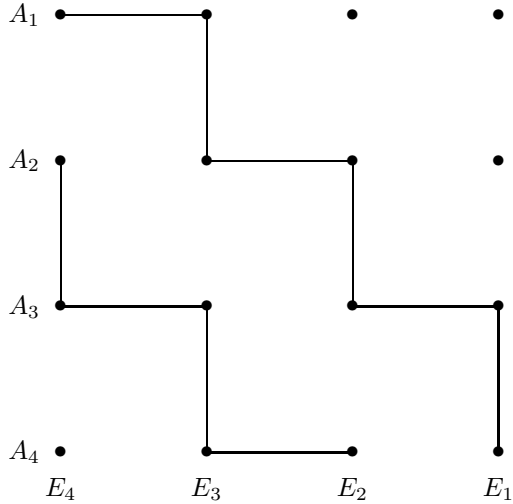
The determinant \eqref{dlth} gives a weighted enumeration of all possible lattice paths satisfying the above constraints, where a configuration with $s$ paths is given a weight of $\exp\haak{\imath\theta s}$.

Families of lattice paths of this type are in bijection with cyclically symmetrical plane partitions \cite{bij,lozenge}. A plane partition of an integer $N$ is an array of integers $n_{j,k}$, such that $n_{j+1,k}\geq n_{j,k}$, $n_{j,k+1}\geq n_{j,k}$ and $N=\sum_{j=1}^{\infty}\sum_{k=1}^{\infty}n_{j,k}$. Plane partitions can be represented by a pile of unit cubes by introducing $x$,$y$,$z$ coordinates in $\mathbb{Z}^{3}$ and placing at position $(j,k,0)$ a stack of $n_{j,k}$ unit cubes. A cyclically symmetric plane partition is a plane partition whose representation as a pile of cubes is symmetric under a cyclic permutation of the x,y,z coordinates. The bijection maps lattice paths of the above type to cyclically symmetric plane partitions that fit in an $L\times L\times L$ box . Families of lattice paths containing $n$ paths are mapped to cyclically symmetric plane partitions with $n$ cubes on the main diagonal.

Numerical calculations indicate that the expansion \eqref{asymp} is also valid for complex $\theta$. It is thus possible to obtain the asymptotics for real values for the weight of a lattice path from this expansion. Assigning a weight of $w$ to a path yields for the asymptotics of the weighted enumeration $N\haak{w}$:
\begin{equation}
N\haak{w}\propto \haak{\frac{3 \sqrt{3}}{4}}^{L^{2}}w^{\frac{L}{2}}L^{\frac{3\log\haak{w}^{2}}{4\pi^{2}}+\frac{7}{36}}
\end{equation}

Another interesting problem is the enumeration of nonintersecting lattice paths with fixed numbers of paths. The number of families with $n$ paths is the coefficient of $\exp\haak{\imath n\theta}$ of the determinant \eqref{dlth}, which is the coefficient $C_{n}\haak{L}$ of the characteristic polynomial of the $L\times L$ Pascal matrix, see \eqref{pasc}. It is possible to derive an asymptotic formula for this number if $n=L/2$. Fourier transforming the asymptotic formula for $D\haak{L,\theta}$ w.r.t.\ $\theta$ by expanding $a_{0,0}\haak{\theta}$ about $\theta=0$ yields:
\begin{equation}
\begin{split}
C_{L/2}\haak{L}=&\sqrt{\frac{\pi}{3}}\haak{\frac{8}{3}}^{\frac{1}{36}}\haak{
\frac{\Gamma\haak{\frac{1}{3}}}{
\Gamma\haak{\frac{2}{3}}}}^{\frac{2}{3}}\exp\haak{\frac{1}{3}\zeta'\haak{-
1}}\haak{\frac{3\sqrt{3}}{4}}^{L^{2}}L
^{\frac{7}{36}}\haak{\log\haak{L}}^{-\frac{1}{2}}\\
&\times\haak{1+O\haak{\frac{1}{\log\haak{L}}}}
\end{split}
\end{equation} 
Here we have used the exact value for $a_{0,0}\haak{0}$, given in Table \ref{amp}.

\section{Probability that a point is on a non-contractible loop}
For even $L$ loops can wind round the cylinder, while for odd $L$ there is a path spanning the length of the cylinder. In \cite{mitradmtcs} we have conjectured that the 
probability, $W\haak{L}$, that a point is on such a non-contractible loop is given as:
\begin{equation}
W\haak{L}=\frac{A\haak{L}}{A_{\text{HT}}\haak{L}^{2}}
\end{equation}

In this section we will show that this follows from the conjectures \eqref{plm} 
and \eqref{plmodd}. Consider the two faces of the lattice next to some point on a loop. The loop can surround either one of the faces or for even $L$ wind round the cylinder, or for odd $L$, it can span the entire cylinder. The probability $W\haak{L}$ is thus the probability that the loop does not surround either one of the two faces. Denote the probability that the face to the left is surrounded by $n$ loops and the face to the right by $m$ loops as $P\haak{L,n,m}$. Since both faces are surrounded by zero loops if and only if the loop does not 
surround any of the faces we have:
\begin{equation}
W\haak{L}=P\haak{L,0,0}
\end{equation}
Either the number of loops surrounding the left face differs by $1$ from the 
number of loops surrounding the right face, or both faces are surrounded by zero 
loops. It thus follows that the probability 
that a single face is surrounded by $m$ loops, $P\haak{L,m}$, can be expressed 
in terms of the $P\haak{L,a,b}$ 
as
\begin{equation}
P\haak{L,m}=P\haak{L,m,m+1}+P\haak{L,m,m-1}
\end{equation}
for $1\leq m < \floor{L/2}$,
\begin{equation}
P\haak{L,0}=P\haak{L,0,1}+P\haak{L,0,0}
\end{equation}
and
\begin{equation}
P\haak{L,\floor{L/2}}=P\haak{L,\floor{L/2},\floor{L/2}-1}
\end{equation}
These equations, together with $P\haak{L,n,m}=P\haak{L,m,n}$, which follows from 
the fact that the loop model 
is symmetric under half turn rotations, imply:
\begin{equation}
W\haak{L}=\sum_{k=0}^{L/2}P\haak{L,k}\haak{-1}^{k}
\end{equation}
According to \eqref{phidef} this is $\gem{\phi\haak{L,-1}}$ and according to 
conjectures \eqref{plm} and \eqref{plmodd} this is given by $\frac{D\haak{L,2\pi/3}}{A_{\text{HT}}\haak{L}^{2}}$ which simplifies to 
$\frac{A\haak{L}}{A_{\text{HT}}\haak{L}^{2}}$ (see Table \ref{eval}).

\section{Some conjectures about probabilities involving connectivities of points 
on a row}
In this section we give a conjecture for the probability that $n$ consecutive 
points on a row are on different 
loops and, for $n\leq 4$, conjectures for the probability that they are on the 
same loop. These conjectures 
are valid for even $L$.

Let $S\haak{L,n}$ be the probability that $n$ consecutive points are on 
different loops. We conjecture that:
\begin{equation}
S\haak{L,n}=\frac{T\haak{L,n}}{T\haak{2 n,n}}\frac{1}{A\haak{n}^2}
\end{equation}

where

\begin{equation}
T\haak{L,n}=\frac{\prod_{m=2}^{n}\prod_{k=1}^{\frac{m}{2}}\rhaak{L^2-4\haak{m-
k}^2}}{\prod_{k=0}^{\frac{n}{2}-
1}\rhaak{L^2-\haak{2 k+1}^2}^{n-1-2k}}
\end{equation}

The limit $L\rightarrow\infty$ yields for large $n$:

\begin{equation}
S\haak{\infty,n}\propto\haak{\frac{64}{27}}^{-\frac{n^2}{2}}n^{\frac{7}{36}}
\end{equation}

Let $R\haak{L,n}$ be the probability that $n$ consecutive points on a row are on 
the same loop. We conjecture that:

\begin{equation}
\begin{split}
R\haak{L,2}=& \frac{11 L^{2} +4}{16\haak{L^2-1}} \\
R\haak{L,3}=& \frac{53 L^{4}+44 L^{2}+128}{128\haak{L^{2}-1}^{2}}\\
R\haak{L,4}=&\frac{5}{2^{15}}\frac{1993 L^{8}-11544 L^{6}-12528 L^{4}-150784 
L^{2}-294912}{\haak{L^{2}-9}\haak{L^{2}-1}^{3}}
\end{split}
\end{equation}

\section{Conclusion}
We have presented new conjectures on exact expressions for correlations in the dense O$(1)$ loop model on $L\times\infty$ square lattices with periodic boundary conditions. We have generalized an earlier obtained result for the probability that a point is surrounded by $m$ loops to odd $L$. Using this conjecture we have calculated the expectation value of the operator that changes the weight of loops winding round a given point. It turns out that the expectation value is given by a binomial determinant that is known to give a weighted counting of cyclically symmetric plane partitions and also of certain types of families of nonintersecting lattice paths. By calculating the asymptotics of the expectation value of this operator, we have obtained the asymptotics of this binomial determinant. For special values of the loop-weights the determinant can be calculated in closed form. For those values we have exactly evaluated the amplitude function of the leading asymptotic behavior of the expectation value of the operator.

Finally we have given conjectures on the probability that $n$ consecutive points are all on different loops and, for $n\leq 4$, that they are on the same loop.

\section{Acknowledgments}
This work was supported by Stichting FOM, which is part of the Dutch foundation of scientific research, NWO. We thank Christian Krattenthaler, Greg Kuperberg and Jan de Gier for useful conversations.

\appendix

\section{Asymptotics of correlations}\label{ap:barnes}
The asymptotic behavior of correlations that are given as simple product 
formulae can be obtained by using 
the Euler-Maclaurin summation formula. However, this approach will yield the 
result up to a constant of 
proportionality. To obtain the results presented in Table \ref{amp}, we need to 
find the asymptotics of 
$A\haak{L}$ and $A_{\text{HT}}\haak{L}$, including the constant term. Both 
$A\haak{L}$ and 
$A_{\text{HT}}\haak{L}$ are given as a product over factorials, and for such 
cases a simple method exists to 
obtain the asymptotic behavior, which is particularly suitable to be implemented in computer algebra systems.

Products over factorials can be expressed in terms of the Barnes G-function \cite{barnes1, barnes2, barnes3, barnsasymp, barnesspec}. This function  
satisfies the relations:
\begin{equation}
\begin{split}
G\haak{z+1}&=\Gamma\haak{z}G\haak{z}\\
G\haak{1}&=1
\end{split}
\end{equation}
Where $\Gamma\haak{z}$ is the gamma function. The Barnes G-function is uniquely defined if the  condition of convexity \cite{conv} is added:
\begin{equation}
\frac{d^3}{dz^3}\log\haak{G\haak{z}}\geq 0 \text{ for } z>0
\end{equation}
In the following we shall write $x!$ for $\Gamma\haak{x+1}$, also for noninteger $x$.
It follows from the definition of the G-function that:
\begin{equation}\label{facg}
\prod_{k=0}^{n}(k+x)!=\frac{G\haak{n+2+x}}{G\haak{1+x}}
\end{equation}
The products occurring in $A\haak{L}$ and $A_{\text{HT}}\haak{L}$ can be 
rewritten as in the above equation by 
using Gauss' multiplication formula for the gamma function. E.g. the numerator 
in $A\haak{L}$, see \eqref{al}, can be rewritten as:
\begin{equation}
\haak{3j+1}!=\prod_{k=1}^{3j+1}k=\prod_{k=1}^{j}3k\prod_{k=0}^{j}\haak{3k+1}
\prod_{k=0}^{j-1}\haak{3k+2}
\end{equation}
By taking out factors of 3 in the products this becomes:
\begin{equation}
\haak{3j+1}!=\frac{3^{3j+1}}{\haak{-\frac{2}{3}}!\haak{-
\frac{1}{3}}!}j!\haak{j+\frac{1}{3}}!\haak{j-\frac{1}{
3}}!
\end{equation}
From \eqref{facg} it thus follows that:
\begin{equation}\label{prodfact}
\prod_{j=0}^{n-1}\haak{3j+1}!=\frac{3^{\frac{3}{2}n^{2}-\half 
n}}{\haak{-\frac{2}{3}}!^{n}\haak{-
\frac{1}{3}}!^{n}}\frac{G\haak{n+1}G\haak{n+\frac{4}{3}}G\haak{n+
\frac{2}{3}}}{G\haak{\frac{4}{3}}G\haak{\frac{2}{3}}}
\end{equation}

To obtain the asymptotic behavior of our correlations, we thus need the 
asymptotics of the Barnes G-function 
and some special values of this function. According to \cite{barnsasymp}, the 
asymptotic behavior of the Barnes G-function is:
\begin{equation}
\begin{split}
\log\haak{G\haak{z+1}}=&z^{2}\haak{\half\log{z}-\frac{3}{4}}+\half\log\haak{2\pi}z-
\frac{1}{12}\log\haak{z}+\zeta'\haak{
-1}\\
\text{}&-\sum_{k=1}^{\infty}\frac{B_{2k+2}}{4k\haak{k+1}z^{2k}}
\end{split}
\end{equation}
Here the $B_{k}$ are the Bernoulli numbers, and $\zeta'\haak{z}$ is the derivative of the zeta function. 

The denominator of \eqref{prodfact} contains the term $G\haak{4/3}G\haak{2/3}$. Such products can be simplified using the multiplication formula for the G-function \cite{multg}:
\begin{equation}
\begin{split}
G\haak{nz}=&\exp\rhaak{\haak{1-n^{2}}\zeta'\haak{-1}}n^{n^{2}z^{2}/2-nz+5/12}\\
&\times\haak{2\pi}^{\frac{n-1}{2}\haak{1-nz}}\prod_{i=0}^{n-1}\prod_{j=0}^{n-1}G\haak{z+\frac{i+j}{n}}
\end{split}
\end{equation}
To express the product $G\haak{4/3}G\haak{2/3}$ in terms of elementary functions one chooses $z=1/3$ and $n=3$.

Proceeding in this way we have obtained:
\begin{equation}
A\haak{L}\sim\haak{\frac{8}{3}}^{\frac{1}{36}}\haak{\frac{\Gamma\haak{\frac{2}{3}}}{\Gamma\haak{\frac{1}{3}}}}^{\frac{1}{3}}\exp\haak{\frac{1}{3}\zeta'\haak{-1}}\haak{\frac{3\sqrt{3}}{4}}^{L^{2}}L^{-\frac{5}{36}}
\end{equation}
And
\begin{equation}\label{hlf}
\begin{split}
A_{\text{HT}}\haak{L}&\sim H\haak{L}\text{ for even }L\\
A_{\text{HT}}\haak{L}&\sim 3^{-\frac{1}{4}}H\haak{L}\text{ for odd }L
\end{split}
\end{equation}
where
\begin{equation}
H\haak{L}=\haak{\frac{4}{3}}^{\frac{1}{18}}\haak{\frac{\Gamma\haak{\frac{1}{3}}}{\Gamma\haak{\frac{2}{3}}}}^{\frac{1}{3}}\exp\haak{\frac{2}{3}\zeta'\haak{-1}}\haak{\frac{3\sqrt{3}}{4}}^{\frac{L^{2}}{2}}L^{\frac{1}{18}}
\end{equation}

\end{document}